\documentclass[11pt,twoside 
]{article}

\usepackage{baaa2009}
\usepackage{graphicx}
\usepackage{subfigure}
\usepackage{psfrag}
\usepackage{amssymb}
\usepackage[spanish,activeacute,english]{babel}
\usepackage[latin1]{inputenc}
\usepackage[T1]{fontenc} % Computer Modern (CM) fonts
\usepackage{ae,aecompl}  % and: dvips -Pcmz -Pamz macros_aaa.dvi
\usepackage{latexsym}
\usepackage{verbatim}
\usepackage{amsmath}
\usepackage{stmaryrd}
\usepackage{amsfonts}
\usepackage{amssymb}
\usepackage{wasysym}
\usepackage[colorlinks=true,dvips]{hyperref}

\begin{document}

%%%%%%%%%%%%%%%%%%%%%%%%%%%%%%%%%%%%%%%%%%%%%%%%%%%%%%%%%%%%%%%%%%%%%%%%%%%
%%%%% SELECCIONE EL IDIOMA EN QUE SE ESCRIBE EL ARTÍCULO:              %%%%
%\myselectspanish
\myselectenglish
%%%%%%%%%%%%%%%%%%%%%%%%%%%%%%%%%%%%%%%%%%%%%%%%%%%%%%%%%%%%%%%%%%%%%%%%%%%

\vskip 1.0cm
\markboth{ N. A. Casco }%
{Surface roughness estimation of a parabolic reflector}

\pagestyle{myheadings}
%%%%% DESCOMENTE LA LINEA QUE DESCRIBE EL CARACTER DE SU TRABAJO       %%%%
%\vspace*{0.5cm}
%%\noindent TRABAJO INVITADO 
%\noindent PRESENTACIÓN ORAL
\noindent PRESENTACIÓN MURAL
%%\noindent RESUMEN 
\vskip 0.3cm
\title{Surface roughness estimation of a parabolic reflector}

%%\title{ Template paper for publication in the Bulletin of the 
%%Argentinian Astronomical Association with instructions for the use of 
%%\LaTeX{}}
%
\author{Nicolás A. Casco}
\affil{Instituto Argentino de Radioastronom\'{\i}a (IAR)}

\begin{abstract}
Random surface deviations in a reflector antenna reduce the aperture efficiency.
This communication presents a method for estimating the mean surface deviation of a 
parabolic reflector from a set of measured points.
The proposed method takes into account systematic measurement errors, such as the offset between the origin of reference frame and the vertex of the surface, and the misalignment between the surface rotation axis and the measurement axis.
The results will be applied to perform corrections to the surface of one of the \mbox{30 m} diameter radiotelescopes at the Instituto Argentino de Radioastronom\'{i}a (IAR).
\end{abstract}

\begin{resumen} 
La rugosidad superficial de una antena reflectora es uno de los par\'{a}metros
que reduce la eficiencia de la apertura. En este trabajo se presenta un m\'{e}todo para
la estimaci\'{o}n de la rugosidad superficial de una antena parab\'{o}lica a
partir de un conjunto de puntos medidos. El m\'{e}todo propuesto corrige
ciertos errores sistem\'{a}ticos de la medici\'{o}n, como la falta de
coincidencia entre el punto de referencia de las mediciones y el v\'{e}rtice
de la superficie, y la desalineaci\'{o}n entre el eje de revoluci\'{o}n de la
superficie y el eje de la medida. Los resultados obtenidos ser\'{a}n aplicados
para realizar correcciones a la superficie de uno de los radiotelescopios de
\mbox{30 m} de di\'{a}metro del Instituto Argentino de Radioastronom\'{i}a (IAR).
\end{resumen}

\vspace{-1em}
\section{Introduction}

Superficial imperfections of a reflector antenna reduce its performance and limits its maximum working frequency. Under certain general assumptions, the Ruze Criterion (Ruze 1966; Zarghamee 1967; Balanis 1982; Baars 2007) allows evaluation of the loss $\alpha$ in the antenna gain for a given wavelength $\lambda$, as a function of \textit{rms} surface error $\varepsilon$, 

\begin{equation}
\alpha  = e^{ - \left( {\frac{{4\pi \varepsilon }}{\lambda }} \right)^2 }
\label{eq:ruze1}
\end{equation}

\begin{figure}[htbp]
	\centering
	\hfill%
	\subfigure[Gain losses $\alpha$ as a function of $\varepsilon/\lambda$ .]{%
	\includegraphics[ width=0.48\textwidth ]{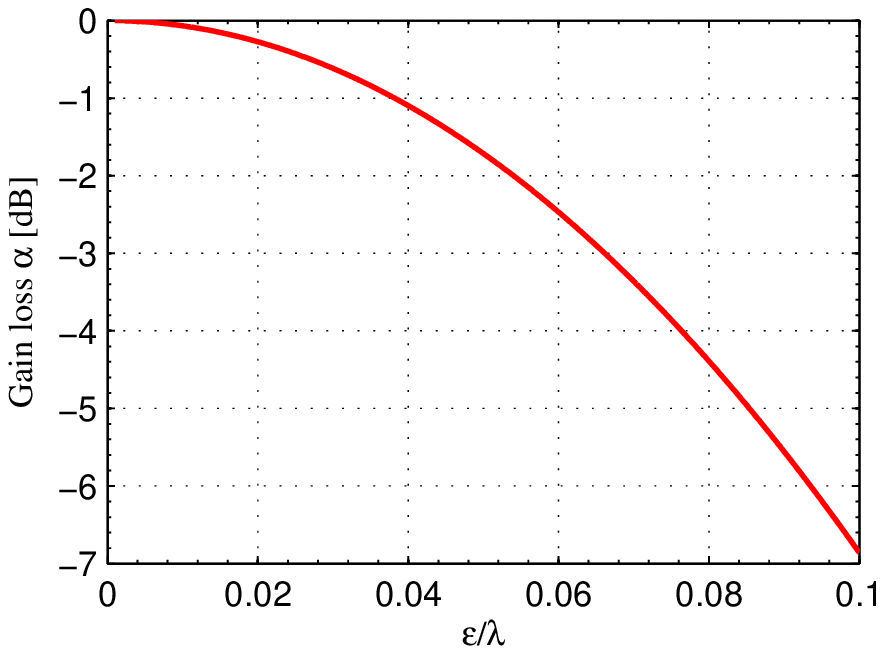}~\hfill
	\label{fig:surface_loss}%
	}%subfigure
	\subfigure[Antenna II at IAR gain as a function of wavelength for different values of the roughness $\varepsilon$.]{%
  \includegraphics[width=0.48\textwidth]{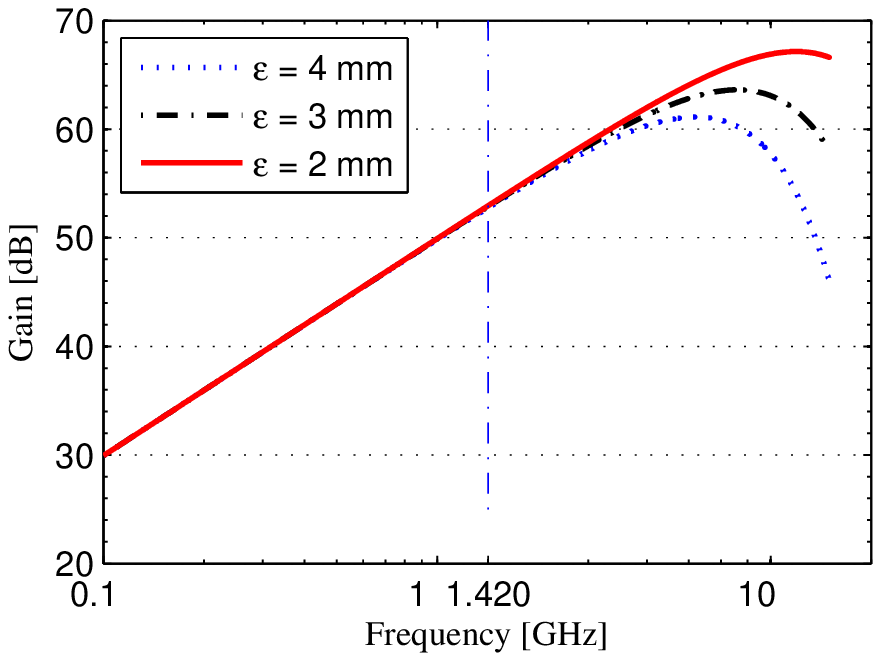}\hfill~%
	\label{fig:gain_loss}%
	}%subfigure
	\label{fig:efectos}
	\caption{Effects of the surface roughness $\varepsilon$ on the gain of a reflector antenna.}
\end{figure}

Figure \ref{fig:surface_loss} shows the reduction in the gain of a reflector antenna as a function of ${\varepsilon}/{\lambda}$. The effects on the gain as a function of wavelength for different values of ${\varepsilon}$ can be seen in Figure \ref{fig:gain_loss}. The plot corresponds to calculations made for a \mbox{30 m} diameter parabolic reflector antenna, like Antenna II at IAR. 

Periodical determination of $\varepsilon$ is required to perform the necessary corrections (Parker \& Srikanth 2001).  In this work we present an algorithm that estimates the surface roughness and other surface parameters from a set of measured points.  

\vspace{-1em}
\section{Fitting algorithm}
The fitting algorithm uses a parametric model of the surface $\bf{X}$ adapted from Ahn (2005), 
\begin{equation}
{\bf{X}}({\bf{a}},{\bf{u}}) = \bf{R}\left( {a_\theta  ,a_\varphi  } \right)
\left[ 
	\left( 
		{\begin{array}{*{20}c}
				{u_1 \cos (u_2)}  \\
				{u_1 \sin (u_2)}  \\
				{\frac{{u_1^2 }}{{4a_f }}} \\
		\end{array}} 
	\right) +
 	\left( 
 		{\begin{array}{*{20}c}
   		{a_x }  \\
   		{a_y }  \\
   		{a_z }  \\
		\end{array}} 
	\right) 
\right]
\label{eq:Parametrica}
\end{equation}
\noindent Here ${\bf{u}} :=  \left( {u_1 ,u_2 } \right)^T$ are the parameters that generate the points on the surface, where $(\cdot)^T$ denotes the transpose matrix.
$\bf{R}\left( {a_\theta  ,a_\varphi  } \right)$ is the rotation matrix that corrects the misalignment between the axis of the paraboloid and the axis of the measurement, and $\left( {{a_x },{a_y },{a_z }} \right)^T$ is a translation that compensates the difference between the origin of the measurement coordinate system and the one of the surface. Parameter $a_f$ is the focal length of the ideal parabola.
	
The  algorithm estimates the parameter vector that defines the surface: ${\bf{a}} := \left( {{a_f },{a_x },{a_y },{a_z },{a_\theta},{a_\varphi}} \right)^T$. Unlike previous works that performed an algebraic fitting of the surface (Muravchik et al. 1990; Ahn 2005), here the mean square of the orthogonal distance $d_{i}$ between the surface and the measured points is minimized. This approach has the advantage of yielding the minimum roughness. Although the approach results in an increased computational load and greater complexity, this should pose no problem for current desktop computers and modern programming languages (Eaton 2002). 

The estimated parameter vector $\mathbf{\hat{a}}$ is obtained from the expression
\begin{equation}
\mathbf{\hat{a}} = \mathop{\arg \min	}\limits_{\mathbf{a} \in \mathbb{R}^k }
\sum\limits_{i = 1}^p {d_i^2(\mathbf{a})}
\label{argmin}
\end{equation} 
\noindent The number of parameters to fit is \mbox{$k=6$} and $p$ is the number of measured points. The optimization problem was solved using a \emph{Quadratic Sequential Programming} method. The value of $d_{i}$ was calculated analytically to further improve the performance of the algorithm, see Casco (2008).

\vspace{-1em}
\section{Method validation}

\begin{figure}[hbtp]
	\vspace{-3em}
	\centering
	\hfill%
	\subfigure[Parameter fit error $\mathbf{a} - \mathbf{\hat{a}}$.]{%
  \includegraphics[width=0.47\textwidth]{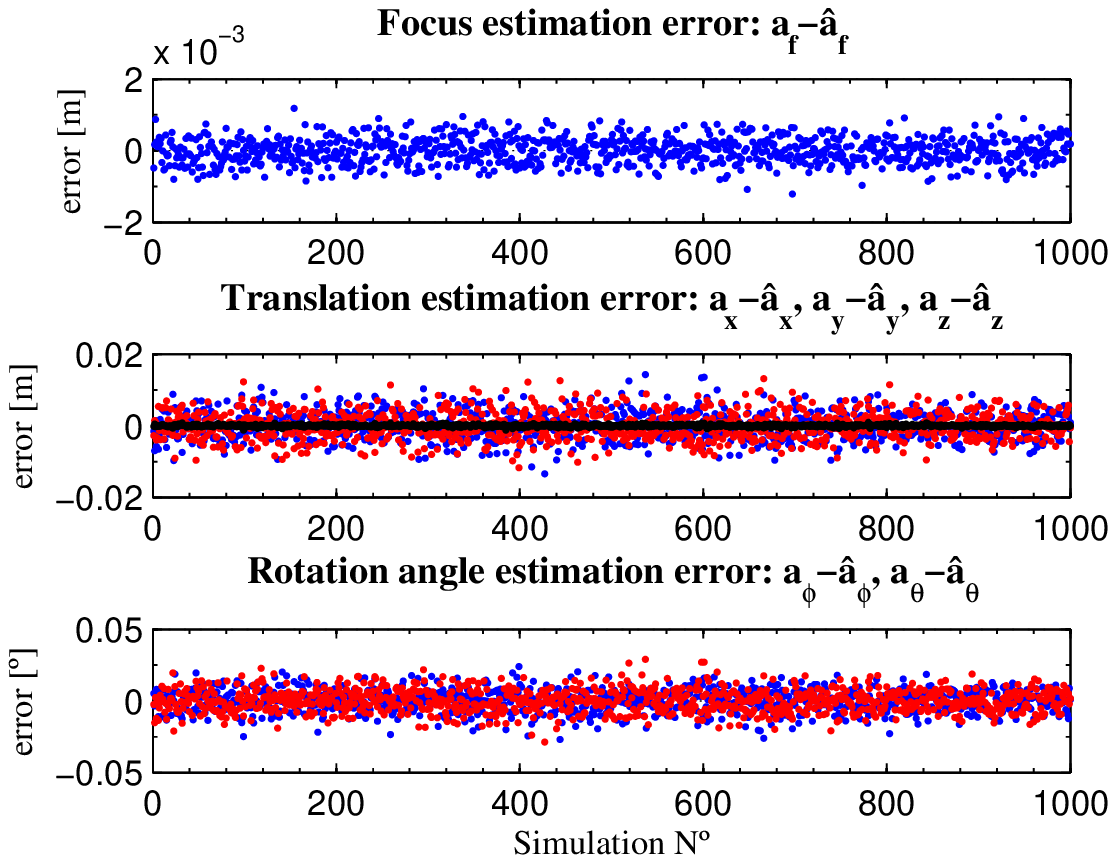}~\hfill
  \label{fig:SimParametros}%
  }%subfigure
  \subfigure[Surface roughness $\varepsilon$ estimation error.]{
  \includegraphics[ width=0.47\textwidth]{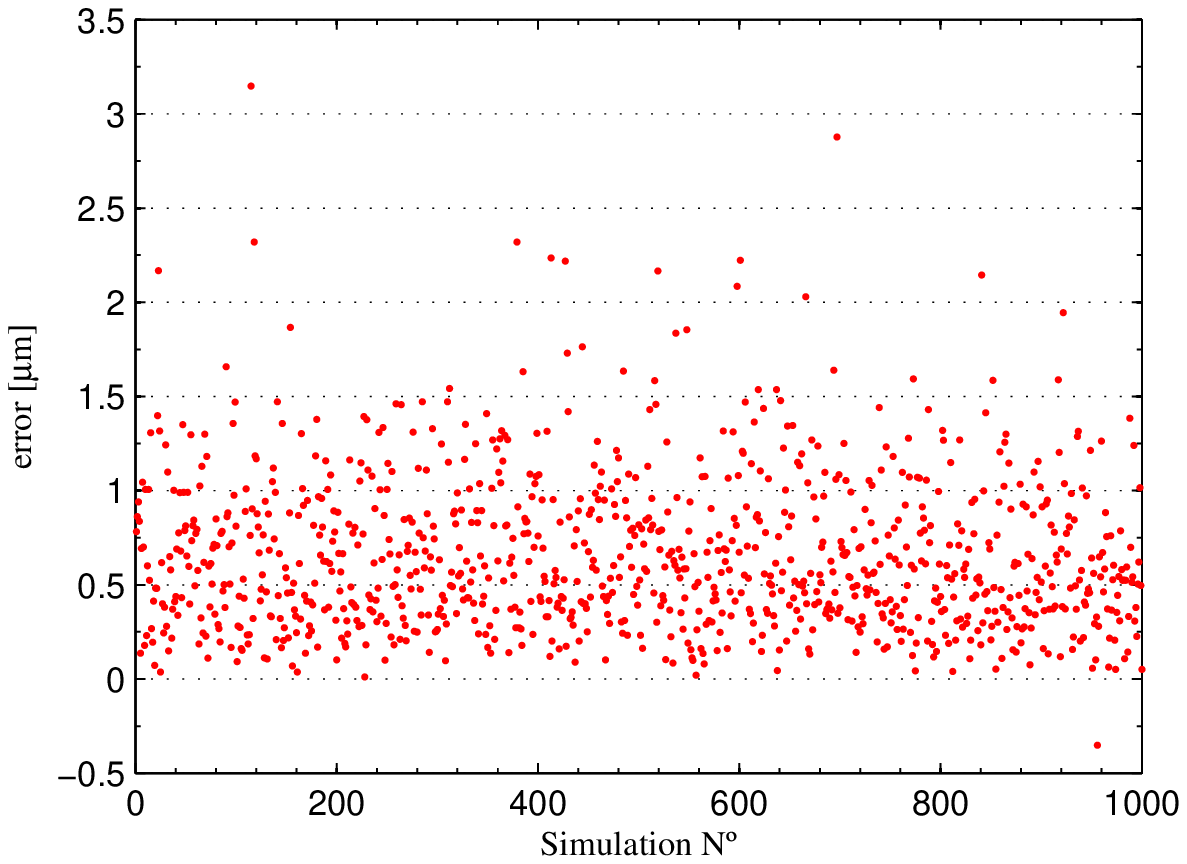}\hfill~%
  \label{fig:SimError}%
  }%subfigure                    
  \caption{Results of the Monte Carlo simulation.}%
  \label{fig:Montecarlo}%
\end{figure}

Monte Carlo simulations (Bevington \& Robinson 2003) were carried out to check the stability of the method and its correct implementation. Each simulation consisted in generating 700 points on a paraboloid of known parameters ${\bf{a}}$, contaminated with measurement noise, and perform the fit to obtain $\mathbf{\hat{a}}$. The position of the synthetically generated points was approximately the same as that of the measured points. The simulation parameters are summarized in \mbox{Table \ref{table:Montecarlo}}.
Figure \ref{fig:Montecarlo} shows the results obtained from a thousand simulation runs using different colours when more than one parameter is plotted on the same graph. It can be concluded that the algorithm is stable and accurate enough for the proposed application.
\begin{table}[htbp]
\renewcommand{\arraystretch}{1.2}
\begin{center}
\tiny{
\begin{tabular}{@{}|c|c|c|}
\hline
\bf{Parameter} & \bf{Variation range} & \bf{Statistical distribution}
\\ %[0.1cm]
\hline 
Focus                               & 12.5$\pm1.5$ m  & Uniform  \\ %[0.1cm]
\hline 
Translations $x$,$y$,$z$            & $10$ cm         & Uniform  \\ %[0.1cm]
\hline 
Rotations                         & $\pm5^{\circ}$    & Uniform  \\ %[0.1cm]
\hline  
\textbf{$\varepsilon$ }& $5$ mm          & Gaussian \\ %[0.1cm]
\hline
\end{tabular}
}
\caption{Monte Carlo simulation parameters.}
\label{table:Montecarlo}
\end{center}
\end{table}

\vspace{-2em}
\section{Data visualization}
\begin{figure}[htbp]
  \centering
  \includegraphics[ width=0.60\textwidth]{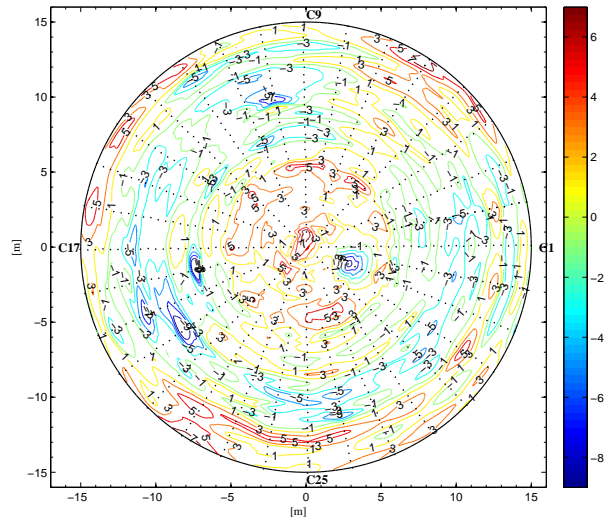}
  \caption{ Contour plots of the surface deformation of the antenna [mm]. Black dots indicate the positions where measurements were made by Cáceres et al. 2008.}
  \label{fig:contour_plot}
\end{figure}

Routines that generate surface roughness contour plots on the antenna, interpolating the processed data, were also developed. This provides a graphical assessment of the results that help to determine possible corrective actions. Figure \ref{fig:contour_plot} is an example of the contour plots obtained.    

\vspace{-1em}
\section{Conclusions}
A processing algorithm for the measurements of the surface of a large reflector antenna was presented. It allows to simultaneously estimate the parameters that define the surface and the systematic measurement errors, in order to minimize the orthogonal distance between the measured points and the ideal surface. The method presents some advantages over previous works (Muravchik et al. 1990), based on an algebraic fitting. The algorithm implementation was validated through Monte Carlo simulations. Furthermore, data visualization routines were developed to ease data assessment. This method will be applied to perform an upgrade to the surface of the Antenna II at IAR. 

%\agradecimientos 
\vspace{-1	em}
\begin{referencias}

\reference S. J. Ahn, \textit{Least Squares Orthogonal Distance Fitting of Curves and Surfaces in Space}, Springer, 2005.

\reference J. W. M. Baars, \textit{The Paraboloidal Reflector Antenna in Radio Astronomy and Comunication. Theory and Practice.}, Springer Science, 2007.

\reference C. A Balanis, \textit{Antenna Theory: Analysis and Design}, John Wiley \& Sons, 1982.

\reference P. R. Bevington, D. K. Robinson, \textit{Data Reduction and Error Analysis for the Physical Sciences}, McGraw-Hill, 3rd Edition, 2003.

\reference J.M. Cáceres, G. M. E. Villa, A. P. Lucchesi, \textit{Determinación de la rugosidad de una superficie}, Facultad de Ingeniería, UNLP, 2008.

\reference N. Casco, ``Estimación de la Rugosidad Superficial de la Antena II'', Informe Interno IAR N$^\circ$ 95, Instituto Argentino de Radioastronomía, 2008.

\reference J. W. Eaton, \textit{GNU Octave Manual}, Network Theory Limited, 2002. 

\reference C. Muravchik, C. Rago, J. A. Bava, A. J. Sanz, ``Método de Verificación de Imperfecciones en Superficies de Antenas Reflectoras Parabólicas'', Informe Interno IAR N$^\circ$ 66, Instituto Argentino de Radioastronomía, 1990.

\reference D.H. Parker, S. Srikanth, ``Measurement system for the Green Bank Telescope'', \textit{IEEE Int. Symposium Antennas and Propagation Society}, vol.4, pp. 592-595, 2001.

\reference J. Ruze, ``Antenna Tolerance Theory A Review'', \textit{Proceedings of the IEEE}, vol. 54, no. 4, pp. 633-640, 1966.

\reference M. S. Zarghamee, ``On Antenna Tolerance Theory'', \textit{IEEE Transactions on Antennas and Propagation}, vol. 15, no. 6, pp. 777-781, 1967.

\end{referencias}

\end{document}